%% file: main.tex
\newcommand{\extended}[1]{}    % LEAVE UNCOMMENTED
\documentclass{article}
\usepackage{ijcai22}

\input{packages.tex}
\usepackage{wojtek15logics,wojtek15logicsWP,wojtek15other,theoremsdecl}
\usepackage{macros}

\newcommand{\state}{\mathit{g}}
\renewcommand{\model}{\mathit{M}}
\newcommand{\AMAS}{S\xspace}
\newcommand{\hatPV}{\mathit{PV}}
\newcommand{\evt}{\alpha}
\newcommand{\evttwo}{\beta}
\newcommand{\events}{\mathit{Evt}}
\newcommand{\roc}{R}

\renewcommand{\lan}[1]{\textbf{#1}\xspace}
\renewcommand{\LTL}{\lan{LTL}}

\renewcommand{\ATL}{\lan{ATL}}
\renewcommand{\LTLX}{\lan{LTL-X}}
\renewcommand{\ATLs}{\ensuremath{\mathbf{ATL^*}}\xspace}
\renewcommand{\sATL}{\ensuremath{\mathbf{sATL}}\xspace}

\renewcommand{\sATLs}{\ensuremath{\mathbf{sATL^*}}\xspace}

\renewcommand{\oneATL}{\ensuremath{\mathbf{1\!ATL}}\xspace}
\renewcommand{\oneATLs}{\ensuremath{\mathbf{1\!ATL^*}}\xspace}

\renewcommand{\ATLir}{\ensuremath{\mathbf{ATL_{ir}}}\xspace}
\renewcommand{\ATLiR}{\ensuremath{\mathbf{ATL_{iR}}}\xspace}
\renewcommand{\AE}{\textbf{AE}\xspace}

\newcommand{\agtchoice}{\textit{\scriptsize E}}
\newcommand{\Std}{\textup{Std}\xspace}
\newcommand{\React}{\textup{React}\xspace}

\newcommand{\strattype}{Y}
\newcommand{\traces}{\mathit{traces}}
\definecolor{tucgreen}{RGB}{0,140,79}

\usetikzlibrary{arrows,automata,calc,shapes,decorations,backgrounds,petri,mindmap,fit,positioning,calc}

\title{Asynchronous Agents with Perfect Recall: Model Reductions, Knowledge-Based Construction, and Model Checking for Coalitional Strategies}

\author{
Dilian Gurov$^{1}$
\and
Filip Jamroga
\and
Wojciech Jamroga$^{2,3}$
\and
Mateusz Kamiński$^{2,3}$
\and
\\
Damian Kurpiewski$^{2,3}$
\and
Wojciech Penczek$^{2}$
\and
Teofil Sidoruk$^{2}$
\affiliations
$^1$KTH Royal Institute of Technology, Stockholm, Sweden\\
$^2$Institute of Computer Science, Polish Academy of Sciences, Warsaw, Poland\\
$^3$Faculty of Mathematics and Computer Science, Nicolaus Copernicus University, Toruń, Poland\\
\emails
\vspace{4px}
dilian@kth.se,
\{w.jamroga, m.kaminski, d.kurpiewski, penczek, t.sidoruk\}@ipipan.waw.pl
}

\begin{document}

\maketitle

\begin{abstract}
Model checking of strategic abilities for agents with memory is a notoriously hard problem, and very few attempts have been made to tackle it. In this paper, we present two important steps towards this goal. First, we take the partial-order reduction scheme that was recently proved to preserve individual and coalitional abilities of \emph{memoryless} agents, and show that it also works for agents with memory. Secondly, we take the Knowledge-Based Subset Construction, that was recently studied for synchronous concurrent games, and adapt it to preserve abilities of memoryful agents in asynchronous MAS.

On the way, we also propose a new execution semantics for strategies in asynchronous MAS, that combines elements of Concurrent Game Structures and Interleaved Interpreted Systems in a natural and intuitive way.
\end{abstract}

%=================================================
\input{intro.tex}
%=================================================
\input{preliminaries.tex}
%=================================================
\input{KBconstruction.tex}
%=================================================
\input{reductions.tex}
%=================================================
\section{Conclusions}\label{sec:conlusions}
%=================================================

Model checking for strategies with imperfect information and perfect recall is well known to be hard, and there have been very few attempts to tackle it in practice. At the same time, it is extremely important, e.g., for the verification of critical systems.
In this paper, we present several steps towards practical verification of the problem.
First, we adapt the Knowledge-Based Subset Construction of~\cite{Gurov22KBC} so that, combined with an \ATLir model checker, it provides a sound but incomplete method of model checking for \iR-strategies.
Secondly, we propose an on-the-fly model reduction scheme that preserves \iR-type abilities.
On the way, we also propose a new execution semantics for strategies in asynchronous MAS, which combines elements of Concurrent Game Structures and Interleaved Interpreted Systems in a natural and intuitive way.

%=================================================
%\section*{Acknowledgments}
%
%  The authors acknowledge the support of the National Centre for Research and Development, Poland (NCBR), and the Luxembourg National Research Fund (FNR), under the PolLux/FNR-CORE project STV (POLLUX-VII/1/2019).
%
%=================================================

\small
\balance

\bibliographystyle{plain}
%\bibliography{ijcai22}
\bibliography{wojtek,wojtek-own}

%=================================================
%\clearpage
%\appendix

%\input{appendix.tex}

\end{document}

%% file: packages.tex
\usepackage{times}
\usepackage{soul}
\usepackage{url}
\usepackage[hidelinks]{hyperref}
\usepackage[utf8]{inputenc}
\usepackage[small]{caption}
\usepackage{graphicx}
\usepackage{amsmath}
\usepackage{amsthm}
\usepackage{booktabs}
\usepackage{algorithm}
\usepackage{algorithmic}
\usepackage{mathrsfs}
\usepackage[T1]{fontenc}
\usepackage{algorithmic}
\usepackage{verbatim}
\usepackage{color}
\usepackage{stmaryrd}
\usepackage{enumerate}
\usepackage{enumitem}
\usepackage{xspace}
\usepackage{balance} % for balancing columns on the final page
\usepackage{multirow}
\usepackage{adjustbox}
\usepackage{amssymb}
\usepackage{cleveref}
\usepackage{tikz}
\usetikzlibrary{decorations,decorations.pathmorphing,arrows,shapes,%
  automata,backgrounds,fit,calc,petri,patterns,matrix}

%% file: intro.tex
%=================================================
\section{Introduction}\label{sec:intro}
%=================================================

Nowadays, multi-agent systems involve a complex, densely connected network of social and technological components.
Such components often exhibit self-interested, goal-directed behavior, which makes it harder to predict and analyze the dynamics of the system.
In consequence, formal specification and automated verification can be of significant help.

\para{Strategic ability.}
Many important properties of multi-agent systems refer to \emph{strategic abilities} of agents and their groups.
For example, functionality requirements can be phrased as the ability of a legitimate agent (or group of agents) to complete their task. Similarly, some security properties refer to the inability of the ``intruder'' to compromise the integrity, secrecy, or anonymity provided by the system.

Two logical formalisms,
\emph{alter\-nating-time temporal logic} \ATLs~\cite{Alur02ATL,Schobbens04ATL} and \emph{Strategy Logic} \SL~\cite{Mogavero14behavioral}, have been proposed for specification and reasoning about strategic properties of MAS.
For instance, the \ATLs formula $\coop{taxi}\,\Always\,\neg\prop{fatality}$ expresses that the autonomous cab can drive in such a way that no one gets ever killed. Similarly, $\coop{taxi,passg}\,\Sometm\,\prop{destination}$ says that the cab and the passenger have a joint strategy to arrive at the destination, no matter what the other agents do.

\para{Verification of strategic ability.}
Specifications in agent logics can be used as input to algorithms and tools for \emph{model checking},
which have been in constant development for over 20 years~\cite{Alur98mocha-cav,Busard15reasoning,Cermak15mcmas-sl-one-goal,Huang14symbolic-epist,Kurpiewski21stv-demo,Lomuscio17mcmas,Kaminski24STVKH}.
Unfortunately, model checking of strategic abilities is hard, both theoretically and in practice.
First, it suffers from the well-known state/transition-space explosion.
Secondly, the space of possible strategies is at least exponential \emph{on top of the state-space explosion}, and incremental synthesis of strategies is not possible in general -- especially in the realistic case of agents with partial observability.
The situation is particularly bad for the variants of \ATLs and \SL~that assume strategies with memory.
Even for the most restricted syntax of ``vanilla'' \ATL, model checking of its imperfect information variants is undecidable for agents with perfect recall~\cite{Dima11undecidable}, and even for coalitions with finite (but unbounded) memory~\cite{Vester13ATL-finite}.
%The theoretical results concur with outcomes of empirical studies on benchmarks~\cite{Busard15reasoning,Jamroga19fixpApprox-aij,Lomuscio17mcmas}, as well as recent attempts at verification of real-life multi-agent scenarios~\cite{Jamroga20Pret-Uppaal,Kurpiewski19embedded}.
On the other hand, the case of memoryful agents is extremely important for the verification of critical systems, where the requirements of integrity and security should hold even against powerful attackers.

\para{Contribution.}
Our novel contributions are as follows:
\begin{itemize}
\item We propose a new class of models for multi-agent interaction (Input/Output Concurrent Game Structures with Imperfect Information, i/o-iCGS in short), and define new, intuitive execution semantics for asynchronous MAS~\cite{Jamroga18por,Jamroga21paradoxes-kr} in terms of such models;

\item We adapt the notions of perfect recall strategies~\cite{Alur02ATL,Schobbens04ATL} and finite recall strategies~\cite{Vester13ATL-finite} to asynchronous MAS;

\item We adapt Partial Order Reductions~\cite{Peled93representatives,Jamroga18por} to model checking of strategies with imperfect information and perfect recall, more precisely the fragment of \ATLiR with no nested strategic modalities;

\item We adapt Knowledge-Based Construction~\cite{Gurov22KBC} to asynchronous MAS, and show how it can be used to provide a sound but incomplete model checking algorithm for the positive fragment of \ATLiR in asynchronous MAS;
%
%\item We report on experimental evaluation of both methods for model checking of simple \ATLiR formulas.
\end{itemize}

\para{Related work.}
Representations and models of asynchronous multi-agent systems follow the idea of APA-nets (asynchronous, parallel automata nets)~\shortcite{Priese83apa-nets}, later extended to interleaved interpreted systems~\cite{lomuscio10partialOrder} and strategic asynchronous MAS with flexible repertoires~\cite{Jamroga18por,Jamroga21paradoxes-kr}. The models proposed in this paper (Section~\ref{sec:io-cgs}) are equally expressive to those in~\cite{Jamroga21paradoxes-kr}, but much more understandable and easy to use.

The Knowledge-based Subset Construction (KBSC) derives from the well-known subset construction for nondeterministic automata~\cite{Rabin59subsetConstruction}. The idea was used further to reduce the verification of 2-player imperfect information games with perfect recall strategies to solving perfect information games with memoryless strategies~\cite{Reif84twoplayergames,Chatterjee07omegareg}, and also for coalitions that are based on instantaneous distributed knowledge~\cite{Guelev11atl-distrknowldge}. 
Our Section~\ref{sec:KBconstruction} extends and adapts the more sophisticated variant~\cite{Gurov22KBC} that was proposed for coalitional abilities in concurrent synchronous games with reachabilty/safety objectives. Our version adapts the construction to \emph{asynchronous multi-player games}, and extends the preservation result to arbitrary \LTL-definable objectives.\footnote{
  In fact, the proofs work for all $\omega$-regular objectives. }

Partial order reductions were proposed for distributed systems to mitigate state-space explosion due to interleaving of independent events~\cite{Peled93representatives,Holzmann94spinPOR,Gerth99por}.
They were further extended to temporal-epistemic properties in multi-agent systems~\cite{lomuscio10partialOrder}, and recently to strategic properties of memoryless agents in asynchronous MAS~\cite{Jamroga18por,Jamroga20POR-JAIR,Jamroga21paradoxes-kr}.
In Section~\ref{sec:por}, we adapt the latter to provide a partial-order reduction scheme for simple formulas of \ATLiR, i.e., the logic for reasoning about (and verification of) strategies with imperfect information and perfect recall.

Model checking of \ATLiR is long known to be undecidable in general~\cite{Dima11undecidable} and \EXPTIME-complete to nonelementary for the special case of singleton coalitions~\cite{Guelev11atl-distrknowldge}. Consequently, there have been very few attempts at practical model checking of the logic. In fact, we are only aware of the (sound but incomplete) approach via bounded model checking in~\cite{Huang15ATLiR}. No implementation is available.

%% file: preliminaries.tex
%\section{Strategic Abilities in Asynchronous MAS}
\section{Models of Asynchronous Interaction}
\label{sec:models}

Models of strategic ability in asynchronous interaction have been inspired by~\cite{lomuscio10partialOrder,Priese83apa-nets} and defined formally in~\cite{Jamroga20POR-JAIR,Jamroga21paradoxes-kr} using \emph{interleaved interpreted systems}. We begin by presenting the representations of asynchronous MAS proposed in~\cite{Jamroga20POR-JAIR,Jamroga21paradoxes-kr}. Then, we propose an equivalent, but more reader friendly formulation of their execution semantics.

\subsection{Asynchronous Multi-Agent Systems}
\label{sec:amas}

To represent the possible behaviors of agents in a MAS, we use modular representations inspired by networks of synchronizing automata~\cite{Priese83apa-nets}, endowed with repertoires of strategic choices.

\begin{definition}[Asynchronous MAS~\cite{Jamroga20POR-JAIR,Jamroga21paradoxes-kr}]\label{def:amas}
An \emph{asynchronous multi-agent system (AMAS)} $\AMAS$ consists of $n$ agents $\A = \set{1,\dots,n}$,
each associated with a tuple $\agent_i =(L_i, \iota_i, \events_i, \roc_i, T_i, \PV_i, V_i)$ including a set
of \emph{local states} $L_i = \{l_i^1, l_i^2,\dots,l_i^{n_i}\}$,
a designated \emph{initial state} $\iota_i \in L_i$,
a nonempty finite set of \emph{events} $\events_i=\{\evt_i^1,\evt_i^2,\ldots, \evt_i^{m_i}\}$,
and a \emph{repertoire of choices}
$\roc_i: L_i \to \powerset{\powerset{\events_i}}$.
For each $l_i\in L_i$, $\roc_i(l_i)=\set{\agtchoice_1,\dots,\agtchoice_m}$ is a nonempty list of nonempty choices available to $i$ at $l_i$.
If the agent chooses $\agtchoice_j = \set{\evt_1,\evt_2,\dots}$, then only an event in $\agtchoice_j$ can be executed at $l_i$ within the agent's module.
Moreover, $T_i \subseteq L_i \times \events_i \times L_i$ is a \emph{local transition relation}, where
$(l_i,\evt,l_i') \in T_i$ represents that event $\evt\in \bigcup\roc_i(l_i)$ changes the local state from $l_i$ to $l_i'$.
Agents are endowed with mutually disjoint, finite and possibly empty sets of \emph{local propositions} $\PV_i$,
and their \emph{valuations} $V_i : L_i \then \powerset{\PV_i}$.
\end{definition}

While each agent ``owns'' the events affecting its state, some of them may be shared with other agents. Such events can only be executed synchronously by all the involved parties. This allows agents to influence the evolution of others' local states.
Moreover, the agent's strategic choices are restricted by its repertoire function.
Assigning sets rather than single events in $\roc_i$, which subsequently determines the type of strategy functions in Section~\ref{sec:atl},
is a deliberate decision, allowing to avoid certain semantic issues that fall outside the scope of this paper.
We refer the reader to~\cite{Jamroga21paradoxes-kr} for the details.

\input{toyexample}

\subsection{Input/Output Concurrent Game Structures}
\label{sec:io-cgs}

To provide an intuitive execution semantics for AMAS, we propose a new class of models, called 
\emph{input/output concurrent game structures with imperfect information (I/O iCGS)}.
I/O iCGS associate an AMAS with a set of global states -- each being a tuple of local states, one per agent -- and a set of global transitions, each labeled by a combination a the agents' choices (the input) and the event that occurs as their consequence (the output). 

Each of their transitions is associated with a label of the form $\inp/\outp$,
where $\inp$ refers to the \emph{choices} made by agents in accordance with their repertoires,
and $\outp$ denotes the \emph{event} executed as a result of these choices.
Note that agents can miscoordinate by making mutually exclusive choices regarding shared events.
We account for such occurrences by means of auxiliary $\epsilon$-loops, where $\epsilon$ is a ``silent'' event that does not belong to any agent.

\begin{definition}[I/O iCGS execution semantics for AMAS]\label{def:ex}
Let $\AMAS$ be an AMAS with $n$ agents $\A = \set{1,\dots,n}$, such that $\epsilon\notin\events_i$ for all $i=1,\dots,n$.
%let $\iota \subseteq \iota_1\times\ldots\times \iota_n$,
Let $Agent(\evt) = \set{i\in\A \mid \evt \in \events_i}$ be the set of agents who have access to {event }$\evt$.
The execution semantics for $\AMAS$ is provided by the I/O iCGS
$\model =(\A, \States, \iota, \events, \PV, V, \{\roc_i\}_{i\in\A}, T, \{\sim_i\}_{i\in\A})$,
where:

\begin{itemize}[leftmargin=*]
\item $\States \subseteq L_1\times\ldots\times L_n$ is the set of \emph{global states},
collecting all the configurations of local states reachable from the \emph{initial global state}
$\iota = (\iota_1,\ldots,\iota_n)$ by $T$ (see below).
We use $\state^i$ to denote agent $i$'s component in $\state$, i.e., $(l_1,\dots,l_n)^i = l_i$;

\item $\events = \bigcup_{i \in \A} \events_i \cup \set{\epsilon}$ is the set of all events, where $\epsilon$ is an auxiliary ``silent'' event with $Agent(\epsilon) = \emptyset$;
\item $\PV = \bigcup_{i\in\A} \PV_i$ is the set of all propositions;
\item $V: \States \rightarrow \powerset{\PV}$ is the global valuation of propositions, defined as $V(g) = \bigcup_{i\in\A} V_i(g^i)$; 

\item $\roc_i: L_i \to \powerset{\powerset{\events_i}}$, for every $i\in\A$, is agent $i$'s repertoire of choices. To simplify the notation, we also extend the repertoires to global states in the obvious way: $\roc_i(g) = \roc_i(g^i)$;

\item $T \subseteq \States \times (\powerset{\events_1}\times\ldots\times\powerset{\events_n}) \times \events \times \States$ is the \emph{global transition relation}, such that
$(\state,\inp,\outp,\state') \in T$ iff $\inp=(\agtchoice_1,\ldots,\agtchoice_n)$,
where $\agtchoice_i \in \roc_i(\state^i)$,
and either:

\noindent \phantom{*}(*)
$\begin{cases}
\outp \in \events \setminus \set{\epsilon},\text{ and}\\
\forall_{i \in Agent(\outp)} \; \outp \in \agtchoice_i \land (\state^i,\outp,(\state')^i) \in T_i,\text{ and}\\
\forall_{i \in \A \setminus Agent(\outp)} \; (\state')^i = \state^i,
\end{cases}$

\noindent \;\;or:\\(**)
$\begin{cases}
\outp = \epsilon,\text{ and}\\
\state' = \state,\text{ and}\\
\forall_{\evt\in Evt\setminus\set{\epsilon}} \; \exists_{i\in Agent(\evt)}\ \evt\notin\agtchoice_i.
\end{cases}$

\item $\sim_i$, for every $i\in\A$, is a relation such that $\state \sim_i \state'$ iff $\state^i = (\state')^i$, i.e., states $\state, \state' \in \States$ are \emph{observationally indistinguishable} for agent $i$.
In other words, $\sim_i$ links states in which agent $i$ makes the same observation, captured by its local state $l_i$. 
\end{itemize}

We will call $M$ the \emph{model of $\AMAS$}, and denote it by $model(\AMAS)$.

In the following, we will write $\state \trns{\inp/\outp} \state'$ instead of $(\state,\inp,\outp,\state') \in T$, and refer to any I/O iCGS defined as above simply as a \emph{model}.
\end{definition}

Intuitively, (**) in the definition of $T$ says that $\epsilon$-loops are added whenever some agents' choices (available in their repertoires $\roc_i$) may potentially lead to a deadlock. Otherwise, as per (*), one of the ``proper'' events will be executed, in which case we follow the corresponding agent(s)' local evolution accordingly.

Note that, without $\outp$, the above modeling structure corresponds to a standard iCGS with synchronous execution according to the tuple $\inp$.
Without $\inp$, it is essentially equivalent to that of an asynchronous Interleaved Interpreted System (IIS), defined as in \cite{Jamroga21paradoxes-kr}.

We complete this section by adapting the definition of an even being \emph{enabled} by a profile of agents' choices, which will later serve to define which execution paths are enabled by a tuple of strategies.

\begin{definition}[Enabled events]\label{def:enabled}

Let $\model$ be a model.
Event $\evt \in \events$ is \emph{enabled} at state $\state\in \States$ iff $\state \trns{\inp/\evt} \state'$ for some tuple of choices $\inp$ and a successor state $\state'$.
The set of such events is denoted by $enabled_\model(\state)$.

Let $A=\set{a_1,\dots,a_k} \subseteq\A$ and $\overrightarrow{\agtchoice_A} = (\agtchoice_{a_1},\dots,\agtchoice_{a_k})$
such that $\agtchoice_i\in R_i(\state^i)$ for every $i\in A$.
Event $\evttwo \in \events$ is \emph{enabled by the vector of choices $\overrightarrow{\agtchoice_A}$ at $\state\in \States$} iff $\state \trns{\inp/\evt} \state'$ for some tuple of choices $\inp$ such that 
$\inp_i = \agtchoice_i$ for $i\in A$, and a successor state $\state'$.
We denote the set of such events by $enabled_\model(\state,\overrightarrow{\agtchoice_A})$.
Clearly, $enabled_\model(\state,\overrightarrow{\agtchoice_A}) \subseteq enabled_\model(\state)$.
\end{definition}

\section{Reasoning About Strategies}\label{sec:atl}

Here, we present the logical formalism that we will use to specify properties of memoryful agents in asynchronous MAS. Our presentation is based on the syntax and semantics of \emph{alternating-time temporal logic \ATLs}, proposed for synchronous MAS in~\cite{Alur97ATL,Alur02ATL,Schobbens04ATL,Vester13ATL-finite}, and adapted to reasoning about memoryless asynchronous agents in~\cite{Jamroga20POR-JAIR,Jamroga21paradoxes-kr}.

\subsection{Strategies}

We consider three types of strategies: memoryless (where the only memory that the agent can use is included in its current local state of the system), perfect recall (where the proponent is allowed to remember the whole history of their observations), and finite memory (where the agent is assumed to have a finite -- though not necessarily bounded -- set of memory states). In all cases, we assume that the agent has imperfect information about the current state of the interaction.

We first cite an established definition of memoryless imperfect information strategy. Then, we propose how to adapt the existing definitions of strategies with memory to reasoning about asynchronous MAS.

\begin{definition}[\ir-strategy~\cite{Schobbens04ATL,Jamroga21paradoxes-kr}]\label{def:memoryless-strategy}
A \emph{memoryless imperfect information strategy} (\ir-strategy) for agent $i$
is defined by a function $\strat_i \colon L_i \to \powerset{\events_i}$,
such that $\strat_i(l) \in \roc_i(l)$ for each $l \in L_i$.
\end{definition}

\begin{definition}[\iR-strategy, adapted from~\cite{Schobbens04ATL}]\label{def:perfect-strategy}
A \emph{perfect recall, imperfect information strategy} (\iR-strategy) for $i$
is defined by $\strat_i \colon L_i^+ \to \powerset{\events_i}$,
such that $\strat_i(h) \in \roc_i(l)$ for each local state $l \in L_i$ and history $h \in L_i^+$. 
For asynchronous systems, we require that $\strat_i$
is \emph{stuttering-invariant}, i.e., if $last(h)=l$ then $\strat_i(hlh') = \strat_i(hh')$.
\end{definition}

\begin{definition}[\iF-strategy, adapted from~\cite{Vester13ATL-finite}]\label{def:finite-strategy}
A \emph{finite-memory, imperfect information strategy} for~$i$ (\iF-strategy) is defined by a \emph{deterministic transducer}:
$$ Z_i = (Q_i, q^0_i, L_i, \events_i, \delta_i, \gamma_i) $$
where $Q_i$ is a finite nonempty set of memory states and $q^0_i\in Q_i$ is the initial memory state. We assume that each memory state $q$ contains sufficient information to identify the current location of agent $i$, denoted $loc_i(q)$. The input alphabet is given by $L_i$ and the output alphabet by $\powerset{\events_i}$, i.e., $Z_i$ takes $i$'s observations as input and returns $i$'s choices as output.
Then, $\delta_i \colon Q_i \times L_i \rightarrow Q_i$ is a stuttering-invariant memory update function, i.e., if $loc_i(q)=l$ then $\delta_i(q,l)=q$. Further, $\gamma_i \colon Q_i \to \powerset{\events_i}$ is an output function that always produces a valid choice, i.e., $\gamma_i (q) \in \roc_i (loc_i(q))$.
Based on the current memory state $q \in Q_i$, the transducer determines the next choice $\strat_i (q)=\gamma_i(q)$ as the current selection of agent~$i$. After a transition, the memory state, initially~$q^0_i$, is updated according to~$\delta_i$ based on the new local state (``observation'') $l' \in L_i$ of~$i$.
\end{definition}

A collective strategy for coalition $A\subseteq\A$ is simply a tuple of individual strategies, one per agent in $A$.

\subsection{Outcome Paths}

Let $\strattype \in \set{\ir,\iF,\iR}$.
We denote the set of $\strattype$-strategies by $\Sigma_i^{\strattype}$.
Note that $\strattype$-strategies are uniform by construction, as they are based on local, and not global states.
Joint strategies $\Sigma_A^{\strattype}$ of coalition $A=\set{a_1,\dots,a_k} \subseteq\A$ are defined as usual,
i.e., as tuples of strategies $\strat_i$, one for each agent $i \in A$.
By $\strat_A(\state) = (\strat_{a_1}(\state),\dots,\strat_{a_k}(\state))$, we denote the joint choice of coalition $A$ at global state $\state$.
An infinite sequence of global states and events $\seq = \state_0 \evt_0 \state_1 \evt_1 \state_2\dots$ is
called a \emph{path} if for every $j \geq 0$ we have $\state_j \trns {\inp_j/\evt_j} \state_{j+1}$
for some choices of agents $\inp_j$.
The $j$-th global state of path $\seq$ is denoted by $\seq[j]$,
and the set of all paths in model $M$ starting at $\state$ by $\Pi_M(\state)$.

\begin{definition}[Standard outcome~\cite{Jamroga21paradoxes-kr}]\label{def:outcome}
Let $A \subseteq \A$, $\strattype \in \set{\ir,\iR}$.
The \emph{standard outcome} of strategy $\strat_A\in\Sigma_A^\strattype$
in state $\state$ of model $\model$
is the set $\outcome_M^\Std(\state,\strat_A) \subseteq \Pi_M(\state)$, such that
$\seq = \state_0 \evt_0 \state_1 \evt_1 \dots \in \outcome_M^\Std(\state,\strat_A)$
iff $\state_0 = \state$,
and for each $m \geq 0$ we have that $\evt_m \in enabled_M(\state_m,\strat_A(\state_m))$.
\end{definition}

\begin{definition}[Reactive outcome~\cite{Jamroga21paradoxes-kr}]\label{def:reactive-outcome}
Let $A \subseteq \A$, $\strattype \in \set{\ir,\iR}$.
The \emph{reactive outcome} of strategy $\strat_A\in\Sigma_A^\strattype$
in state $\state$ of model $\model$
is the set $\outcome_M^\React(\state,\strat_A) \subseteq \outcome_M^\Std(\state,\strat_A)$, such that
$\seq = \state_0 \evt_0 \state_1 \evt_1 \dots \in \outcome_M^\React(\state,\strat_A)$
iff $\evt_m = \epsilon$ implies $enabled_M(\state_m,\strat_A(\state_m)) = \set{\epsilon}$.
\end{definition}
Intuitively, the standard outcome collects all the paths where agents in $A$ follow $\strat_A$, while the others freely choose from their repertoires.
The reactive outcome includes only those outcome paths where the opponents cannot miscoordinate on shared events.
Note that in reactive outcomes of \ir-strategies, $\epsilon$-loops can only occur at the end of paths (i.e., as infinite suffixes). Since executions where opponents' choices lead to a deadlock are excluded by \Cref{def:reactive-outcome}, $\epsilon$-loops can only appear as a result of a miscoordination by the coalition agents, and the memoryless semantics prevents them from making different choices in the same local states.

On the other hand, standard \ir-outcomes may also include interleavings of  $\epsilon$-loops and ``proper'' events, because in this setting the opponents freely choose from their repertoires, not being bound by any strategy. Thus, in particular, they may eventually choose to synchronise after a finite number of miscoordinations with the coalition.

Under the perfect recall semantics, both types of paths can occur in standard as well as reactive outcomes.

\subsection{Syntax of \texorpdfstring{\ATLs}{ATL*}}

Let $\PV$ be a set of propositions
and $\A$ the set of all agents.
The syntax of \emph{alternating-time logic}  is given by:
\begin{center}
$\varphi::= \prop{p} \mid \neg \varphi \mid \varphi\wedge\varphi \mid \coop{A}\gamma$, \hspace{1cm}
$\gamma::=\varphi \mid \neg\gamma \mid \gamma\land\gamma \mid \Next\gamma \mid \gamma\Until\gamma$, \\
\end{center}
where $\propp \in \PV$, $A \subseteq \A$, $\Next$ stands for ``next'',
$\Until$ for ``until'',
and $\coop{A}\gamma$ for ``agent coalition $A$ has a strategy to enforce $\gamma$''.
Boolean operators and temporal operators $\Release$ (``release''), $\Sometm$ (``eventually''), and $\Always$ (``always'') are defined as usual.

It is often beneficial (from the computational complexity point of view) to consider the restricted syntax of ``vanilla'' \ATL, where every occurrence of a strategic modality is immediately followed
by a single temporal operator.
In that case, ``release'' is not definable from ``until'' anymore~\cite{Laroussinie08expATL}, and must be added explicitly to the syntax.
Formally, the language of \ATL is defined by the following grammar:
\begin{center}
$\varphi ::= \prop{p}  \mid  \lnot \varphi  \mid  \varphi\land\varphi  \mid
  \coop{A} \Next\varphi  \mid  \coop{A} \varphi\Until\varphi  \mid \coop{A} \varphi\Release\varphi$.
\end{center}

In the rest of this paper, we are mainly interested in formulas that do not use the next step operator $\Next$, and do not contain nested strategic modalities.
We denote the corresponding subsets of \ATLs and \ATL by \sATLs (``simple \ATLs'') and \sATL (``simple \ATL'').
Moreover, \oneATLs is the fragment of \sATLs that admits only formulas consisting of a single strategic modality
followed by an \LTL formula (i.e., $\coop{A}\gamma$, where $\gamma\in\LTL$), and analogously for \oneATL.

\subsection{Semantics}

The semantics of \ATLs and its fragments, parameterized by the strategy type $\strattype$ and the outcome type $x$, is defined as follows. 

\begin{definition}[Semantics of \ATLs]\label{def:semantics}
Let $x\in\set{\Std,\React}$ and $\strattype\in\set{\ir,\iR}$.
The semantics of \ATLs, for $\strattype$-strategies and $x$-outcomes, is defined by the clauses~\cite{Alur02ATL,Schobbens04ATL,Jamroga21paradoxes-kr}:
\begin{description2}
\setlength\itemindent{-0.25cm}
\item[{$\model,\state \satisf[\strattype]^x \prop{p}$}]
    iff $\prop{p} \in V(\state)$, for $\prop{p} \in \PV$;
\item[{$\model,\state \satisf[\strattype]^x \neg\varphi$}]
    iff $\model,\state \not\satisf[\strattype]^x  \varphi$;
\item[{$\model,\state \satisf[\strattype]^x \varphi_1\land\varphi_2$}]
    iff $\model,\state \satisf[\strattype]^x \varphi_1$ and $\model,\state \satisf[Y] \varphi_2$;
\item[{$\model,\state \satisf[\strattype]^x \coop{A}\gamma$}]
    iff there is a strategy $\strat_A\!\in\!\Sigma_A^\strattype$
    such that for all $\seq\!\in\! \outcome_\model^x(\state,\strat_A)$
    we have $\model,\seq\! \satisf[\strattype]^x\! \gamma$.

\item[{$\model,\seq \satisf[\strattype]^x \varphi$}]
    iff $\model,\seq[0]\satisf[\strattype]^x \varphi$;
\item[{$\model,\seq \satisf[\strattype]^x \neg\gamma$}]
    iff $\model,\seq \not\satisf[\strattype]^x \gamma$;
\item[{$\model,\seq \satisf[\strattype]^x \gamma_1\land\gamma_2$}]
    iff  $\model,\seq\satisf[Y] \gamma_1$ and $\model,\seq \satisf[\strattype]^x \gamma_2$;
\item[{$\model,\seq \satisf[\strattype]^x \Next\gamma$}]
    iff $\model,\seq[1,\infty] \satisf[\strattype]^x \gamma$;
\item[{$\model,\seq \satisf[\strattype]^x \gamma_1\Until\gamma_2$}]
    iff $\model,\seq[i,\infty]\ \satisf[\strattype]^x\ \gamma_2$ for some $i\ge 0$
    and $\model, \seq[j,\infty] \satisf[\strattype]^x \gamma_1$ for all $0\leq j< i$.
\end{description2}
\end{definition}

Additionally, we define 
$\model \satisf[\strattype]^x \varphi$ iff $\model,\iota \satisf[\strattype]^x \varphi$, and 
$\AMAS \satisf[\strattype]^x \varphi$ iff $model(\AMAS) \satisf[\strattype]^x \varphi$.

%% file: toyexample.tex
\newcommand{\voterno}{1}

\begin{figure}[t]\centering
\begin{minipage}{.45\textwidth}\centering
\begin{tikzpicture}[->,>=stealth',shorten >=1pt,auto,node distance=2.1cm,transform shape,semithick,scale=0.8]
       \tikzstyle{every state}=[fill=none,draw=black,text=black,minimum size=1.25cm]

  \node[state] (s0) {$q_0^\voterno$}; %[label={[label distance=0.0cm]90:Voter $i$}]
  \node[state] (s1) [below left=1cm and 1.5cm of s0]  {$q_a^\voterno$};
  \node[state] (s2) [below right=1cm and 1.5cm of s0] {$q_b^\voterno$};
	\node[state] (s3) [below left of=s1]  {$q_{a,g}^\voterno$};
  \node[state] (s4) [below right of=s1] {$q_{a,n}^\voterno$};
	\node[state] (s5) [below left of=s2]  {$q_{b,g}^\voterno$};
  \node[state] (s6) [below right of=s2] {$q_{b,n}^\voterno$};
	\node[state] (s7) [label=below:{$\prop{voted_{\voterno,a}}$}, below of=s3] {$q_{a,g,s}^\voterno$};
	\node[state] (s8) [label=below:{$\prop{voted_{\voterno,a}}$}, below of=s4] {$q_{a,n,s}^\voterno$};
	\node[state] (s9) [label=below:{$\prop{voted_{\voterno,b}}$}, below of=s5] {$q_{b,g,s}^\voterno$};
	\node[state] (s10) [label=below:{$\prop{voted_{\voterno,b}}$}, below of=s6] {$q_{b,n,s}^\voterno$};
	
  \path (s0)	edge         					node [sloped, anchor=center, above] {$vote_{\voterno,a}$} (s1)
							edge        					node [sloped, anchor=center, above] {$vote_{i,b}$} (s2)
				(s1)	edge         					node [sloped, anchor=center, above] {$gv_{\voterno,a}$} (s3)
							edge        					node [sloped, anchor=center, above] {$ng_\voterno$} (s4)
				(s2)	edge         					node [sloped, anchor=center, above] {$gv_{\voterno,b}$} (s5)
							edge        					node [sloped, anchor=center, above] {$ng_\voterno$} (s6)
        (s3)	edge [bend left=40]		node [above left]{$revote_\voterno$} (s0)
							edge 									node [right] {$stop_\voterno$} (s7)
        (s4)	edge									node [sloped, anchor=center, above] {$revote_\voterno$} (s0)
							edge 									node [left]{$stop_\voterno$} (s8)
        (s5)	edge 									node [sloped, anchor=center, above] {$revote_\voterno$} (s0)
							edge 									node [right] {$stop_\voterno$} (s9)
        (s6)	edge [bend right=40]	node [above right] {$revote_\voterno$} (s0)
							edge 									node [left] {$stop_\voterno$} (s10)
				(s7)	edge [loop right] node {} (s7)
				(s8)	edge [loop left] node {} (s8)
				(s9)	edge [loop right] node {} (s9)
				(s10)	edge [loop left] node {} (s10);

    \end{tikzpicture}
\end{minipage}
\hspace{.08\textwidth}
\begin{minipage}{.45\textwidth}\centering
\begin{tikzpicture}[->,>=stealth',shorten >=1pt,auto,node distance=2.1cm,transform shape,semithick,scale=0.8]
       \tikzstyle{every state}=[fill=none,draw=black,text=black,minimum size=1.25cm]

  \node[state] (s0) {$q_0^c$}; %[label={[label distance=3.1cm]90:Coercer}]  	
  \node[state] (s1) [label=below:{$\prop{revealed_{1,a}}$}] at (180:3.0cm)  {$q_{g,1,a}^c$};
  \node[state] (s2) [label=below:{$\prop{revealed_{1,b}}$}] at (240:3.0cm) {$q_{g,1,b}^c$};
  \node[state] (s3) at (120:3.0cm)	{$q_{n,1}^c$};
  \node[state] (s4) [label=below:{$\prop{revealed_{2,a}}$}] at (300:3.0cm)	{$q_{g,2,a}^c$};
	\node[state] (s5) [label=below:{$\prop{revealed_{2,b}}$}] at (0:3.0cm)	{$q_{g,2,b}^c$};
	\node[state] (s6) at (60:3.0cm)	{$q_{n,2}^c$};

  \path (s0)	edge [bend left=15] node [sloped, anchor=center, below] {$gv_{1,a}$} (s1)
							edge [bend left=15] node [sloped, anchor=center, below] {$gv_{1,b}$} (s2)
							edge [bend left=15]	node [sloped, anchor=center, below] {$ng_1$} (s3)
							edge [bend left=15] node [sloped, anchor=center, above] {$gv_{2,a}$} (s4)
							edge [bend left=15] node [sloped, anchor=center, above] {$gv_{2,b}$} (s5)
							edge [bend left=15]	node [sloped, anchor=center, above] {$ng_2$} (s6)
				(s1)	edge [bend left=15] node [sloped, anchor=center, above] {$return$} (s0)
				(s2)	edge [bend left=15] node [sloped, anchor=center, above] {$return$} (s0)
				(s3)	edge [bend left=15] node [sloped, anchor=center, above] {$return$} (s0)
				(s4)	edge [bend left=15] node [sloped, anchor=center, below] {$return$} (s0)
				(s5)	edge [bend left=15] node [sloped, anchor=center, below] {$return$} (s0)
				(s6)	edge [bend left=15] node [sloped, anchor=center, below] {$return$} (s0);
    \end{tikzpicture}
\end{minipage}
	\caption{ASVR$^2_2$: agents Voter$_1$ (top) and Coercer (down).}
	\label{fig:asvr}
\end{figure}
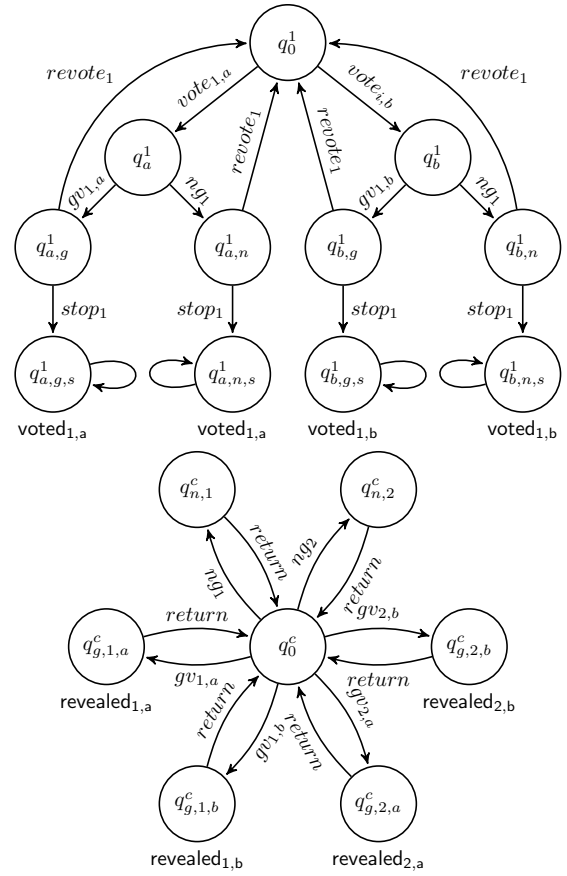

\begin{example}[Asynchronous Simple Voting with Revoting]\label{ex:asvr}
Consider a simple voting system ASVR$_n^k$ with $n+1$ agents ($n$ voters and $1$ coercer) and $k$ candidates.
Each Voter$_i$ agent can cast her vote for a candidate $\set{1,\dots,k}$, and decide whether to share her vote receipt with the Coercer agent. Then she can repeat her voting and vote sharing process.
A graphical representation of the agents for $n=2, k=2$ is shown in Fig.~\ref{fig:asvr}.
The set of propositional variables is $\PV = \set{\prop{voted_{1,1}},\dots,\prop{voted_{n,k}},\prop{revealed_{1,1}},\dots,\prop{revealed_{n,k}}}$.
Proposition $\prop{voted_{i,j}}$ denotes that $v_i$ voted for the $j$-th candidate, while $\prop{revealed_{i,j}}$ denotes
that $v_i$ additionally gave the coercer the proof of having voted for $j$.
The repertoire of the coercer is defined as
$\roc_c(q_0^c)=\set{\set{gv_{1,a},gv_{1,b},ng_{1}},\set{gv_{2,a},gv_{2,b},ng_{2}}}$ and
$\roc_c(q_{g,i,j}^c)=\roc_c(q_{n,i}^c)=\set{\set{return}}$ for $i=1,2$ and $j=a,b$,
i.e., the coercer first \emph{receives} the voter's decision regarding the receipt, and then returns to his previous state.
Analogously, the voter's repertoire is given by: 
$\roc_\voterno(q_0^\voterno)=\set{\set{vote_{\voterno,a}},\set{vote_{\voterno,b}}}$,
$\roc_\voterno(q_j^\voterno)=\set{\set{gv_{\voterno,j}},\set{ng_{\voterno}}}$ 
and
$\roc_\voterno(q_{j,g}^\voterno)=\roc_\voterno(q_{j,n}^\voterno)=\set{\set{revote_\voterno},\set{stop_\voterno}}$ for $j=a,b$.
\end{example}

%% file: KBconstruction.tex
\section{Knowledge-Based Subset Construction for Asynchronous Agents}\label{sec:KBconstruction}

In this section, we adapt the knowledge-based subset construction of~\cite{Gurov22KBC} from synchronous safety/reachability games to asynchronous MAS. Then, we show how it can be used to provide a sound but incomplete model checking algorithm for the one-strategy fragment of \ATLiR~\cite{Schobbens04ATL}.

\subsection{Projection}

Our starting point is an AMAS~$S$ with $n$ agents $\A = \set{1,\dots,n}$, and the I/O iCGS~$\model$ it induces.
The construction consists of two steps.

The first step is a \emph{projection} of~$\model$ 
onto each agent~$i$, resulting in $n$ 
tuples~$\model_i$. 
The projection essentially filters out the elements
in~$\model$ that do not pertain to agent~$i$ (but
retains the global states). In particular, since transitions are labeled by descriptions of the transition ``input'' and ``output,'' the projection retains $i$'s part of both, and removes the other ingredients of transition labels. 

\begin{definition}[Projection]\label{def:projection}
Let $S$ be an AMAS with components as in Definition~\ref{def:amas}, and $\model$ be the induced I/O iCGS with components as in Definition~\ref{def:ex}.
The \emph{projection} of~$\model$ on agent $i \in \A$ is a tuple:
$$\model_i =(\States, \iota, \events_i, \PV_i, V'_i, \roc_i, T'_i, \sim_i)$$
where: 
\begin{itemize}[leftmargin=*]
\item 
$V'_i: \States \rightarrow \powerset{\PV_i}$ 
is defined by
$V'_i (\state) = V_i (\state^i)$.
Note that then $\state \sim_i \state'$ entails $V'_i (\state) = V'_i (\state')$.
\item 
$T'_i \subseteq \States \times \powerset{\events_i} \times (\events_i \cup \set{\epsilon}) \times \States$ 
is defined so that:
\begin{itemize}
\item 
$(\state, \agtchoice_i, \alpha, \state') \in T'_i$ if $(\state, (\agtchoice_1,\ldots,\agtchoice_n), \alpha, \state') \in T$ such that $\alpha \in \agtchoice_i$;

\item 
$(\state, \agtchoice_i, \epsilon, \state') \in T'_i$ if $(\state, (\agtchoice_1,\ldots,\agtchoice_n), \alpha, \state') \in T$ such that $\alpha \not\in \agtchoice_i$.
Note that this includes the case when $\alpha = \epsilon$.
Also note that $(\state, \agtchoice_i, \epsilon, \state') \in T'_i$ only if $\state \sim_i \state'$, and then $V'_i (\state) = V'_i (\state')$.
\end{itemize}
\end{itemize}
\end{definition}

\subsection{Expansion}

The second step of the construction is the application
of a modified version of the standard Knowledge-based Subset Construction~\cite{Reif84twoplayergames,Chatterjee07omegareg} to each projected model~$\model_i$,
resulting in $n$ individual \emph{expansions}~$\model_i^K$.
The idea is to use subsets of~$\States$ to represent possible ``knowledge states'' $s^K$ of agent~$i$. A knowledge state~$s^K$ captures what the agent can deduce about the possible current state of the system. Note that, for asynchronous systems, each $s^K$ must be closed under transitions that are invisible to the agent, i.e., $\epsilon$-transitions.\footnote{
  Similarly to the $\epsilon$-closure in the determinisation of $\epsilon$-NFAs~\cite{Rabin59subsetConstruction}.} 
Thus, the construction starts with the $\epsilon$-closure of the initial state, and keeps adding knowledge states and transitions by looking at the agent's choices and then splitting the possible successors into observational equivalence classes. 

\begin{definition}[Expansion]\label{def:expansion}
Let $S$ be an AMAS with components as in Definition~\ref{def:amas}, $\model$ be the induced I/O iCGS with components as in Definition~\ref{def:ex}, and $\model_i$ be its projection on agent $i \in \A$ with components as in Definition~\ref{def:projection}. 

The \emph{expansion} of~$\model_i$ is a tuple:
$$ \agent_i^K = (\States_i^K, \iota_i^K, \events_i, \roc_i^K, T_i^K, \PV_i, V_i^K) $$
where $\States_i^K$, $\iota_i^K$, and $T_i^K$ are constructed by the procedure below. 
Note that by construction, in every local knowledge state~$s^K \in \States_i^K$, all global states $\state \in s^K$ agree on the local state~$\state^i \in L_i$; we will use $\mathit{loc}_i (s^K)$ to denote that local state. 
Consequently, we can fix
$R_i^K (s^K) = R_i (\mathit{loc}_i (s^K))$ and $V_i^K (s^K) = V_i (\mathit{loc}_i (s^K))$ for every newly added knowledge state~$S^K$.

\begin{enumerate}
\item Let $\iota^K_i = \epsclosure(\{\iota\})$, $\States_i^K = \{\iota^K\}$, and $T_i^K = \emptyset$;

\item For every state $s^K\in\States_i^K$ and choice $\agtchoice\in R_i^K (s^K)$:
    \begin{enumerate}
    \item Take $Succ$ to be the set of $q'\in\States$ such that for some $q\in s^K$ there is a transition $(q,\agtchoice,\evt,q')\in T'_i$ for some $\evt\in\events_i$;

    \item Let $Succ' = \epsclosure(Succ)$;

    \item Let $Succ^K = Succ' / \!\!\!\sim_i$, i.e., the partitioning of $\epsclosure(Succ)$ with respect to the equivalence relation~$\sim_i$;

    \item Add the elements of $Succ^K$ to $\States_i^K$;

    \item\label{exp-step}
    For every $s_{succ}^K\in Succ^K$: if there exists $q\in s^K$, $q'\in s_{succ}^K$, and $\evt\in\events_i$ such that $(q,\agtchoice,\evt,q') \in T'_i$, add $(s^K,\evt,s_{succ}^K)$ to~$T_i^K$;
    \end{enumerate}

\item Continue until no new states have been added to $\States_i^K$.
\end{enumerate}
\end{definition}

% Similar to the standard Subset Construction of Automata Theory, the transition relation in our construction is also deterministic, in the following sense. 

The collection~$\AMAS^K$ of the $n$ individual expansions~$\agent_i^K$ is an AMAS with $n$ agents $\A = \set{1,\dots,n}$.  We refer to it as the \emph{expanded game}.

\subsection{Characterisation Results}

Let $\model^K$ denote the I/O iCGS with components as induced by the AMAS~$\AMAS^K$ in Definition~\ref{def:ex}. 
Let $\state^K$ range over its global states, $\inp$ over tuples $(\agtchoice_1,\ldots,\agtchoice_n)$ of individual choices, and let $\mathit{loc} (\state^K)$ denote the tuple $(\mathit{loc}_1 ((\state^K)^1), \ldots, \mathit{loc}_n ((\state^K)^n))$. 

\begin{lemma}\label{lem:intersection}
For any $\state^K$ in $\model^K$, it holds that $\bigcap \state^K = \set{\mathit{loc} (\state^K)}$.
\end{lemma}

\begin{lemma}[Simulation]\label{lem:global-determinisation}
Let $\AMAS$ be an AMAS, $\model$ be its I/O iCGS, 
$\AMAS^K$ be the expansion of~$\AMAS$, and $\model^K$ its respective I/O iCGS.
Then, for every infinite sequence
$\state_0, \inp_0, \state_1, \inp_1, \state_2, \ldots$
in~$\model$ such that $\iota = \state_0$ and for every $k \geq 0$,
$(\state_k, \inp_k, \outp_k, \state_{k+1}) \in T$ for some $\outp_k$,
there is an infinite sequence
$\state^K_0, \inp_0, \state^K_1, \inp_1, \state^K_2, \ldots$
in~$\model^K$ such that $\iota^K = \state^K_0$ and for every $k \geq 0$,
$\mathit{loc} (\state^K_k) = \state_k$ and
$(\state^K_k, \inp_k, \outp_k, \state^K_{k+1}) \in T^K$.
\end{lemma}

\begin{proof}
Let $\simeq\ \subseteq \States \times \States^K$ be a binary relation defined so that $\state \simeq \state^K$ whenever $\state = \mathit{loc} (\state^K)$.
We show that~$\simeq$ is a \emph{simulation} relation, in the following sense: for $\state \in \States, \state^K \in \States^K$ such that $\state \simeq \state^K$, $(\state,\inp,\outp,\state') \in T$, and $\inp$, $\outp$, and $\state' \in \States$ in~$\model$, then $(\state^K,\inp,\outp,\state'^K) \in T^K$ for some $\state'^K \in \States ^K$ in $\model^K$ such that $\state' \simeq \state'^K$.

Let $\state \in \States$ and $\state^K \in \States^K$ be such that $\state \simeq \state^K$. 
Let $(\state,\inp,\outp,\state') \in T$ for some $\inp$, $\outp$, and $\state' \in \States$ in~$\model$. 
We show that then $(\state^K,\inp,\outp,\state'^K) \in T^K$ for some $\state'^K \in \States ^K$ in $\model^K$ such that $\state' \simeq \state'^K$. 

By Definition~\ref{def:ex}, $\inp = (\agtchoice_1,\ldots,\agtchoice_n)$ for some $\agtchoice_1,\ldots,\agtchoice_n$ such that $\agtchoice_i \in \roc_i(\state)$ for all $i \in \A$,
and either:

\noindent \phantom{*}(*)
$\begin{cases}
\outp \in \events \setminus \set{\epsilon},\text{ and}\\
\forall_{i \in Agent(\outp)} \; \outp \in \agtchoice_i \land (\state^i,\outp,(\state')^i) \in T_i,\text{ and}\\
\forall_{i \in \A \setminus Agent(\outp)} \; (\state')^i = \state^i,
\end{cases}$

\noindent \;\;or:\\(**)
$\begin{cases}
\outp = \epsilon,\text{ and}\\
\state' = \state,\text{ and}\\
\forall_{\evt\in Evt\setminus\set{\epsilon}} \; \exists_{i\in Agent(\evt)}\ \evt\notin\agtchoice_i.
\end{cases}$

\noindent
Since $\agtchoice_i \in \roc_i(\state)$ for all $i \in \A$, and since $\state \simeq \state^K$, by the definition of~$\simeq$ and by Definition~\ref{def:expansion}, 
$R_i^K (\state^K) = R_i (\mathit{loc} (\state^K)) = R_i (\state)$, 
and therefore $\agtchoice_i \in R_i^K (\state^K)$ for all $i \in \A$ in~$\model^K$. 
The proof now proceeds by considering the above two cases in turn. \\

\emph{Case (*).} 
Consider any $i \in Agent(\outp)$. We have that $\outp \in \agtchoice_i$, and by Definition~\ref{def:projection}, we have $(\state,\agtchoice_i,\outp,\state') \in T'_i$ in projection $\model_i$. 

Now, since $\state \simeq \state^K$, by Lemma~\ref{lem:intersection} we have that $\state \in (\state^K)^i$. 

Then, by Definition~\ref{def:expansion}, step~$2 (e)$, there exists $s_i^K \in \States_i^K$ such that $\state' \in s_i^K$ and $((\state^K)^i, \outp, s_i^K) \in T_i^K$ in expansion $\agent_i^K$.

Thus, condition~(*) of Definition~\ref{def:ex} applies to~$\model^K$ with $\state^K$, $\inp$, $\outp$, and with~$\state'^K$ defined by:
$$ (\state'^K)^i = 
\begin{cases}
s_i^K\; \text{~~~~~~if } i \in Agent(\outp) \\
(\state^K)^i \text{~~if } i \in \A \setminus Agent(\outp)
\end{cases} $$
since then the following conditions are satisfied: 
\begin{itemize}
\item[$(i)$]
$\outp \in \events \setminus \set{\epsilon}$
\item[$(ii)$]
$\forall_{i \in Agent(\outp)} \; \outp \in \agtchoice_i \land ((\state^K)^i, \outp, s_i^K) \in T^K_i$
\item[$(iii)$]
$\forall_{i \in \A \setminus Agent(\outp)} \; (\state'^K)^i = (\state^K)^i$
\end{itemize}
and hence, $(\state^K,\inp,\outp,\state'^K) \in T^K$ in~$\model^K$. 

Finally, we have $(\state')^i = \mathit{loc}_i ((\state'^K)^i)$ for all $i \in \A$, and hence $\state' \simeq \state'^K$. \\

\emph{Case (**)}. 
In this case, condition~(**) of Definition~\ref{def:ex} applies to~$\model^K$ with $\state^K$, $\inp$, $\outp = \epsilon$ and $\state'^K = \state^K$, and hence, $(\state^K,\inp,\outp,\state'^K) \in T^K$ in~$\model^K$. Obviously in this case $\state' \simeq \state'^K$ holds.

To conclude the proof of the lemma, it suffices to observe that $\iota \simeq \iota^K$.
\end{proof}

We now introduce some additional notation for observations and their traces. In the definitions below, $\state$ represents a global state in the original model~$M$, $Q$ is a set of global states in~$M$ that may appear as a state in the knowledge-based model $M^K$, $X$~corresponds to either $\state$ or~$Q$; $\pi$~is a single path in either $M$ or~$M^K$, and $\Pi$ is a set of such paths.
\begin{eqnarray*}
obs_i(\state) &=& \state^i \in L_i \\
obs_i(Q) &=& l_i\text{ if }\state^i=l_i\text{ for all }\state\in Q,\\
  && \text{ else undefined} \\
obs_A(X) &=& (obs_i(X) \mid i\in A) \\
\mathit{trace}_A(\state_0\evt_0\state_1\evt_1\dots) &=& (obs_A(\state_0)\, obs_A(\state_1)\, \dots) \\
\traces_A(\Pi) &=& \{\mathit{trace}_A(\pi) \mid \pi\in\Pi  \}
\end{eqnarray*}

That is, an observation of agents $A\subseteq\A$ in state $\state$ is simply a tuple of the corresponding local states, and this is further lifted to sets, sequences, and sets of sequences of states.
We similarly lift $\mathit{loc}$ to traces and sets of traces.

\begin{theorem}\label{thm:preservation}
Let $A\subseteq\A$ be a coalition.
For every joint \ir-strategy~$\strat^K_A$ of $A$ in the expanded game~$\AMAS^K$, there is a joint \iF-strategy~$\strat_A$ in the original game~$\AMAS$, such that:
$$ \traces_\A (\outcome_M^\Std(\iota,\strat_A)) \subseteq
   \mathit{loc} (\traces_\A (\outcome_{M^K}^\Std(\iota^K,\strat^K_A))) $$
\end{theorem}
\begin{proof}[Proof sketch]
Let~$\strat^K_A$ be a joint \ir-strategy in~$\AMAS^K$. 
% We define a joint iR-strategy~$\strat_A$ in~$S$ as follows.
For every agent~$i \in A$, we use $\agent_i^K$ and $\strat^K_i \colon \States_i^K \to \powerset{\events_i}$ to define the transducer:
$$ Z_i = (\States_i^K, \iota_i^K, L_i, \events_i, \delta_i^K, \strat^K_i) $$
with memory states~$\States_i^K$, initial memory state~$\iota_i^K$, input alphabet~$L_i$, output alphabet~$\powerset{\events_i}$, memory update function~$\delta_i^K \colon \States_i^K \times L_i \rightharpoonup \States_i^K$, and output function~$\strat^K_i$, where $\delta_i^K (s_1^K, l_2)$ is defined as the local knowledge state~$s_{succ}^K \in \States_i^K$ described in steps~2(a-d) of Definition~\ref{def:expansion} for the local knowledge state~$s_1^K$, choice $\agtchoice = \strat^K_i (s_1^K)$, and local state $l_2 = \mathit{loc}_i (s_{succ}^K)$, when it exists.

The result is then a direct consequence of Lemma~\ref{lem:global-determinisation}, since following in the original game the \iF-strategy~$\strat_A$, defined by the tuple of transducers~$Z_i$, will yield in the expanded game a simulating path as described by the lemma.
\end{proof}

\input{preservation}

%% file: preservation.tex
\subsection{Preservation of Strategic Abilities}

Theorem~\ref{thm:preservation} says that the outcome of a memoryless strategy $\strat^K_A$ in the knowledge-based expansion $\AMAS^K$ simulates the outcome of its corresponding finite-memory strategy $\strat_A$ in the original AMAS $\AMAS$. In consequence, one can use model checking for memoryless strategies in the knowledge-based model to synthesize knowledge-based strategies in the ``memoryless'' model.
Formally, we have the following.

\begin{theorem}\label{thm:preservation-oneatl-iF}
Let $\AMAS$ be an AMAS and $\varphi$ a formula of $\oneATLs$.
If $\AMAS^K \models_\ir^\Std \varphi$ then $\AMAS \models_\iF^\Std \varphi$.
\end{theorem}
\begin{proof}
Take $\varphi \equiv \coop{A}\gamma$. 
If $\AMAS^K \models_\ir^\Std \varphi$, then there exists an \ir-strategy~$\strat^K_A$ in $\AMAS^K$ such that 
$model(\AMAS^K),\iota \models_\iF^\Std \gamma$ for every $\pi\in\outcome_{model(\AMAS)^K}^\Std(\iota^K,\strat^K_A)$.
By Theorem~\ref{thm:preservation}, there must also be an \iF-strategy~$\strat_A$ in $\AMAS$ such that
$\traces_\A (\outcome_{model(\AMAS)}^\Std(\iota,\strat_A)) \subseteq
   \mathit{loc} (\traces_\A (\outcome_{model(\AMAS)^K}^\Std(\iota^K,\strat^K_A)))$.

Note that, in AMAS, $trace_\A(\pi) = \pi$ for every path $\pi$. This is because observations are local states, and an $\A$-tuple of
local states is the global state. 
Thus, we get that 
$\outcome_{model(\AMAS)}^\Std(\iota,\strat_A) \subseteq \mathit{loc}(\outcome_{model(\AMAS)^K}^\Std(\iota^K,\strat^K_A))$, 
and hence $model(\AMAS),\iota \models_\iF^\Std \gamma$ for every $\pi\in\outcome_{model(\AMAS)}^\Std(\iota^K,\strat_A)$.
So, $model(\AMAS),\iota \models_\iF^\Std \coop{A}\gamma$.
\end{proof}

\begin{theorem}\label{thm:preservation-oneatl-iR}
Let $\AMAS$ be an AMAS and $\varphi$ a formula of $\oneATLs$.
If $\AMAS^K \models_\ir^\Std \varphi$ then $\AMAS \models_\iR^\Std \varphi$
\end{theorem}
\begin{proof}
Straightforward, by observing that the transducer in Theorem~\ref{thm:preservation} implements an \iR-strategy for coalition $A$.
\end{proof}

We conjecture that analogous characterizations hold for the reactive semantics of asynchronous strategic ability, but the actual formalization and proofs are left for future work.

The above results are important because they provide decidable approximations for verification of $\ATL_\iR$ and $\ATL_\iF$ -- both of which are undecidable for even very simple input formulas~\cite{Dima11undecidable,Vester13ATL-finite}. Combined with model checking tools for memoryless imperfect information strategies (e.g.,~\cite{Kurpiewski21stv-demo,Kaminski24STVKH}), they can be used to obtain a sound (though incomplete) algorithm to model-check strategic abilities of agents with memory, which is extremely important for the verification of critical systems. For example, checking for cyberattacks against cryptographic protocols must assume a very powerful attacker, in particular one with high memory capacity.

%% file: reductions.tex
\section{Model Reductions for iR-Strategies}\label{sec:por}

\emph{Partial order reduction (POR)} has been defined for temporal logics~\cite{Peled93representatives,Gerth99por,lomuscio10partialOrder},
and recently extended to reasoning about memoryless strategies with imperfect information~\cite{Jamroga18por,Jamroga20POR-JAIR,Jamroga21paradoxes-kr}.
The idea is to take a set of asynchronous modules (AMAS in our case), and use depth-first search through the space of global states to generate a reduced model that satisfies exactly the same formulas as the full model.
POR strives to remove paths that change only the interleaving order of an ``irrelevant'' event with another event.
Importantly, the method generates the reduced model directly from the representation, without generating the full model at all.

In this section, we adapt the construction of~\cite{Jamroga21paradoxes-kr} to strategies with imperfect information and \emph{perfect recall}.
Note that in this approach, model reductions are always obtained within the context of some formula,
and in particular, of the considered agent coalition.
Hence, in the following, we assume a subset of agents $A \subseteq \A$ whose events are always visible (cf.~\Cref{sec:independence}).

\subsection{Properties of Submodels}\label{sec:submodel}

We begin by formally defining a submodel,
and provide auxiliary lemmas that will be used to prove the correctness of reduction.
Then, we recall the notions of invisibility and independence of events, and of stuttering equivalence.

\begin{definition}[Submodel]\label{def:reducedM}
Consider two models $\model$ and $\model'$.
$\model' = (\A', \States', \events', \PV', V', \roc', T', \sim_i')$ is a \emph{submodel},
or a \emph{reduced model}, of $\model = (\A, \States, \events, \PV, V, \roc, T, \sim_i)$,
denoted by $\model' \subseteq \model$, if we have:
$\A' = \A$,
$\States' \subseteq \States$,
$\events' = \events$,
$\PV' \subseteq \PV$,
$V' = V|_{\States'}$,
$T$ is an extension of $T'$,
$\roc_i' = \roc_i$,
and $\sim_i' = \sim_i$ for all $i\in\A=\A'$.
\end{definition}

\begin{lemma}\label{prop:submodel}
Let $\model' \subseteq \model$, $x \in \set{\Std,\React}$.
Then, for each $\strat_A \in \Sigma_A^{\iR}$, we have
$\outcome_{\model'}^{x}(\iota,\strat_A) = \outcome_{\model}^{x}(\iota,\strat_A) \cap \Pi_{\model'}(\iota)$.
\end{lemma}
\begin{proof}
Note that each joint \iR-strategy in $\model$ is also a well defined joint \iR-strategy in $\model'$
since it is defined on the local states of each agent of an AMAS, which is extended
by both $\model$ and $\model'$.
The lemma follows directly from the definitions of standard and reactive outcomes
(\Cref{def:outcome,def:reactive-outcome}), plus the fact that
$\Pi_{\model'}(\iota) \subseteq \Pi_{\model}(\iota)$.
\end{proof}

\begin{lemma}\label{prop:same-strategy}
Let $\model$ be a model, $x \in \set{\Std,\React}$, $\seq, \seq' \in \Pi_\model(\iota)$,
and for some $i \in \A:$
$\events(\seq)\mid_{\events_i} = \events(\seq')\mid_{\events_i}$.
Then, for each \iR-strategy $\strat_i$, we have
$\seq \in \outcome_{\model}^{x}(\iota,\strat_i)$ iff $\seq' \in \outcome_{\model}^{x}(\iota,\strat_i)$.
\end{lemma}
\begin{proof}
Let $\events(\pi)\mid_{\events_i} = b_0b_1\ldots$ be the sequence of the events of agent $i$ in $\pi$.
It can be finite or infinite.
For each $b_j$ of this sequence let $\pi[b_j]$ denote the global state from which $b_j$ is executed in $\pi$.
By induction we can show that for each $j \geq 0$ such that $b_j$ is defined,
we have $\pi[b_j]^i= \pi'[b_j]^i$.
For $j = 0$ it is easy to see that $\pi[b_0]^i = \pi[b_0]^i = \iota^i$.
Assume that the thesis holds for $j = k$.
The induction step follows from the fact the local evolution $T_i$ can be defined as a (partial) function \cite{Jamroga21paradoxes-kr},
so if $\pi[b_k]^i = \pi'[b_k]^i = l$,
then $\pi[b_{k+1}]^i = \pi'[b_{k+1}]^i = l'$ such that $(l,b_k,l') \in T_i$, for some $l,l' \in L_i$.
This means that for each $j \geq 0 $ such that $b_j$ is defined, we have
$\pi[b_0]^i..\pi[b_j]^i = \pi'[b_0]^i..\pi'[b_k]^i$,
so the local histories $h_i,h'_i \in L^+_i$ are the same in $\pi$ and $\pi'$
at each global state from which an event of agent $i$ is executed.
Thus, by \Cref{def:outcome,def:reactive-outcome}, for each \iR-strategy $\strat_i$
we have $\pi \in \outcome_{\model}^{x}(\iota,\strat_i)$
iff $\pi' \in \outcome_{\model}^{x}(\iota,\strat_i)$, which concludes the proof.
\end{proof}

The lemma can be easily generalized to joint strategies $\strat_A\in\Sigma_A^{\iR}$.
Note that the same property does {not} hold for perfect information strategies.
While the current local state $l_i$ can only change through the execution of an event by agent $i$,
the current global state can possibly change because of another agent's transition.
Similarly, the analogue of \Cref{prop:same-strategy} does not hold in synchronous models of MAS,
since the local transitions of $i$ in a synchronous model can be influenced by the events
selected by the other agents.

\subsection{Independence of Events}\label{sec:independence}

Intuitively, an event is invisible iff it does not change the valuations of the propositions.\footnote{
  This concept of invisibility is technical, and is not connected to the view of any agent\extended{ in the sense of~\cite{MalvoneMS17}}. }
Additionally, we can designate a subset of agents $A$ whose events are visible by definition.
Furthermore, two events are independent iff they are not events of the same agent and at least one of them is invisible.

\begin{definition}[Invisible events]
Consider a model $M$, a subset of agents $A\subseteq\A$, and a subset of propositions $\hatPV\subseteq \PV$.
An event $\evt \in \events$ is {\em invisible}
wrt.~$A$ and $\hatPV$ if $Agent(\evt)\cap A = \emptyset$ and for each two global states
$\state, \state' \in \States$ and any $\inp$, we have that $\state \trns{\inp/\evt} \state'$ implies
$V(\state)\cap\hatPV = V(\state')\cap\hatPV$.
The set of all invisible events for $A,\hatPV$ is denoted by $Invis_{A,\hatPV}$,
and its closure -- of visible events -- by $Vis_{A,\hatPV} = \events \setminus Invis_{A,\hatPV}$.
\end{definition}

\begin{definition}[Independent events]
The notion of \emph{independence} $I_{A,\hatPV}\subseteq \events\times \events$ is defined as:
$I_{A,\hatPV} = \{(\evt,\evt') \in \events \times \events \mid Agent(\evt) \cap Agent(\evt') = \emptyset\}\ \setminus\ (Vis_{A,\hatPV} \times Vis_{A,\hatPV})$.
Events $\evt, \evt' \in \events$ are called {\em dependent} if $(\evt,\evt') \not \in I_{A,\hatPV}$.
If it is clear from the context, we omit the subscript $\hatPV$.
\end{definition}

\subsection{Stuttering Equivalence}\label{sec:stuttering-ltlx}
Let $\model$ be a model, $\model' \subseteq \model$, and $\hatPV \subseteq \PV$ a subset of propositions.
Stuttering equivalence says that two paths can be divided into corresponding finite segments,
each satisfying exactly the same propositions.
Stuttering path equivalence requires two models to always have corresponding, stuttering-equivalent paths.
\Cref{equivs} connects the latter to \LTLX.

\begin{definition}[Stuttering equivalence]
\label{def-ste}
Paths $\seq \in \Pi_{\model}(\iota)$ and $\seq' \in \Pi_{\model'}(\iota)$ are \emph{stuttering equivalent},
denoted $\seq \equiv_{s} \seq'$, if there exists a partition
$B_0 = (\seq[0],\dots,\seq[i_1-1]),\ B_1=(\seq[i_1],\dots,\seq[i_2-1]),\ \ldots$\ of the states of $\seq$,
and an analogous partition $B'_0, B'_1, \ldots$ of the states of $\seq'$,
such that for each $j \geq 0:$ $B_j$ and $B'_j$ are nonempty and finite, and
$V(\state)\cap\hatPV = V'(\state')\cap\hatPV$ for every $\state\in B_j$ and $\state'\in B'_j$.

Models $\model$ and $\model'$ are \emph{stuttering path equivalent},
denoted $\model \equiv_{s} \model'$ if for each path $\seq \in \Pi_{\model}(\iota)$,
there is a path $\seq' \in \Pi_{\model'}(\iota)$ such that $\seq \equiv_{s} \seq'$.\footnote{Typically,
the definition contains also the symmetric condition which in our case always holds for $\model$ and its submodel $\model'$,
as $\Pi_{\model'}(\iota) \subseteq \Pi_{\model}(\iota)$. }
\end{definition}

\begin{theorem}[\cite{cgp99}]\label{equivs}
If $\model \equiv_{s} \model'$, then we have
$\model, \iota \models \varphi$ iff $\model', \iota' \models \varphi$,
for any \LTLX\ formula $\varphi$ over $\hatPV$.
\end{theorem}

\subsection{Preserving Equivalence in Reduced Models}\label{sec:POR-correctness}

Let $x\in\set{\Std,\React}$.
Rather than generating the full model $\model$, one can generate a reduced model $\model'$
satisfying the following property, denoted by $\AE_A^x$:

\vspace{0.2cm}
\noindent
$\forall \seq \in \outcome_{\model}^{x}(\iota,\strat_A)
\exists \seq' \in \outcome_{\model'}^{x}(\iota,\strat_A)
\seq \equiv_s \seq'$, $\forall \sigma_A\!\in\!\Sigma_A^\iR$.
\vspace{0.2cm}    

We first show algorithms that generate reduced model satisfying $\AE_A^x$,
and then prove that these reduced models preserve a subset of \ATLs that excludes nested strategic modalities (which we refer to as \sATLs) for \iR-strategies.

\para{Algorithms for partial order reduction.}\label{sec:PORA}
POR is used to reduce the size of models while preserving satisfaction for a class of formulas.
The standard DFS~\cite{GKPP99} or DDFS~\cite{CVWY92} is modified in such a way
that from each visited state $\state$ an event $\evt$ to compute the successor state $\state_1$
such that $\state \trns{\inp/\evt} \state_1$, is selected from $E(\state)\cup\set{\epsilon}$
such that $E(\state) \subseteq enabled(\state)\setminus\set{\epsilon}$.
That is, the algorithm always selects $\epsilon$, plus a subset of the enabled events at $\state$.
Let $A \subseteq \A$.
The conditions on the heuristic selection of $E(\state)$ given below are inspired by~\cite{peled-on_the_fly,cgp99,Jamroga18por}.
\begin{description}
\item[{\bf C1}]
    Along each path $\pi$ in $\model$ that starts at $\state$,
    each event that is dependent on an event in $E(\state)$
    cannot be executed in $\pi$ without an event in $E(\state)$ being executed first in $\pi$.
	Formally, $\forall \pi \in \Pi_{\model}(\state)$ such that
        $\pi = \state_0\evt_0\state_1\evt_1\ldots$ with $\state_0 = \state$,
	and $\forall b \in \events$ such that $(b,c) \notin I_A$ for some $c \in E(\state)$,
	if $\evt_i = b$ for some $i \geq 0$, then $\evt_j \in E(\state)$ for some $j < i$.
\item[{\bf C2}]
    If $E(\state) \neq enabled(\state)\setminus\set{\epsilon}$, then $E(\state) \subseteq Invis_A$.
\item[{\bf C3}]
    For every cycle in $\model'$ containing no $\epsilon$-transitions,
    there is at least one node $\state$ in the cycle for which
    $E(\state) = enabled(\state)\setminus\set{\epsilon}$,
	i.e., all successors of $\state$ are expanded.
\end{description}

\begin{theorem}\label{prop:stequ}
Let $A \subseteq \A$, $x \in \set{\Std,\React}$, $\model$ be a model,
and $\model' \subseteq \model$ be the reduced model generated by DFS
with the choice of $E(\state')$ for $\state' \in \States'$
given by conditions {\bf C1, C2, C3} and the independence relation $I_A$.
Then, $\model'$ satisfies $\AE_A^x$.
\end{theorem}
\begin{proof}
Notice that the reduction of $\model$ under the conditions {\bf C1, C2, C3} above is equivalent
to the reduction of $\model$ without the $\epsilon$-loops under the conditions
{\bf C1, C2, C3} of~\cite{peled-on_the_fly},
and then adding the $\epsilon$-loops to each state of the reduced model
where there exists a miscoordinating combination of agents' choices.
Although the setting is slightly different, it can be shown similarly to~\cite[Theorem~12]{cgp99}
that the conditions {\bf C1, C2, C3} guarantee that the models:
(i) $\model$ without $\epsilon$-loops and (ii) $\model'$ without $\epsilon$-loops are stuttering path equivalent.
More precisely, for each path $\pi = \state_0\evt_0\state_1\evt_1\cdots$ with $\state_0 = \iota$ (without $\epsilon$-transitions) in $\model$
there is a stuttering equivalent path $\pi' = \state'_0\evt'_0\state'_1\evt'_1\cdots$ with $\state'_0 = \iota$ (without $\epsilon$-transitions) in $\model'$ such that
$\events(\pi)|_{Vis_A} = \events(\pi')|_{Vis_A}$,
i.e., $\pi$ and $\pi'$ have the same maximal sequence of visible events for $A$. \textbf{(*)}

We will now prove that this implies $\model \equiv_s \model'$.
Removing the $\epsilon$-loops from $\model$ eliminates two kinds of paths:
(a) paths with infinitely many ``proper'' events, and
(b) paths ending with an infinite sequence of $\epsilon$-transitions.
Consider a path $\seq$ of type (a) from $\model$.
Notice that the path $\seq_1$, obtained by removing the $\epsilon$-transitions from $\seq$,
is stuttering-equivalent to $\seq$.
Moreover,  by~\textbf{(*)}, there exists a path $\seq_2$ in $\model'$ without $\epsilon$-transitions,
which is stuttering-equivalent to $\seq_1$.
By transitivity of the stuttering equivalence, we have that $\seq_2$ is stuttering equivalent to $\seq$.
Since $\seq_2$ must also be a path in $\model'$, this concludes this part of the proof.

Consider a path $\seq$ of type (b) from $\model$, i.e., $\seq$ ends with an infinite sequence of $\epsilon$-transitions.
Let $\seq_1$ be the sequence obtained from $\seq$ after removing $\epsilon$-transitions,
and $\seq_2$ be any infinite path without $\epsilon$-transitions such that $\seq_1$ is its prefix.
Then, it follows from~\textbf{(*)} that there is a stuttering equivalent path
$\seq_2' = g'_0a'_0g'_1a'_1\cdots$ with $g'_0 = \iota$ in $\model'$ such that $\events(\seq_2)|_{Vis_A} = \events(\seq_2')|_{Vis_A}$.
Consider the minimal finite prefix $\seq_1'$ of $\seq_2'$ such that  $\events(\seq_1')|_{Vis_A} = \events(\seq_1)|_{Vis_A}$.
Clearly, $\seq_1'$ is a sequence in $\model'$ and can be extended with an infinite number of $\epsilon$-transitions to the path $\seq'$ in $\model'$.
It is easy to see that $\seq$ and $\seq'$ are stuttering equivalent.

So far, we have shown that our reduction under the conditions {\bf C1, C2, C3} guarantees that the models
$\model$ and $\model'$ are stuttering path equivalent, and more precisely
that for each path $\pi = \state_0\evt_0\state_1\evt_1\cdots$ with $\state_0 = \iota$ in $\model$
there is a stuttering equivalent path $\pi' = \state'_0\evt'_0\state'_1\evt'_1\cdots$ with $\state'_0 = \iota$ in $\model'$ such that
$\events(\pi)|_{Vis_A} = \events(\pi')|_{Vis_A}$,
i.e., $\pi$ and $\pi'$ have the same maximal sequence of visible events for $A$.
To show that $\model'$ satisfies $\AE_A^x$, consider an \iR-joint strategy $\strat_A$ and $\seq \in \outcome_{\model}^{x}(\iota,\strat_A)$.
As demonstrated above, there is $\seq' \in \Pi_{\model'}(\iota)$ such that $\seq \equiv_s \seq'$
and $\events(\seq)|_{Vis_A} = \events(\seq')|_{Vis_A}$.
Since $\events_i \subseteq Vis_A$ for each $i \in A$, the same sequence of events of each $\events_i$ is
executed in $\seq$ and $\seq'$.
Note that opponent reactiveness only restricts the outcome sets, and not the model itself;
hence, the above reasoning applies to $x = \Std$ as well as $x = \React$.
By the generalization of Lemma~\ref{prop:same-strategy} to \iR-joint strategies we get
$\seq' \in \outcome_{\model}^{x}(\iota,\strat_A)$.
Thus, by \Cref{prop:submodel} we have $\seq' \in \outcome_{\model'}^{x}(\iota,\strat_A)$.
\end{proof}

We now show that the reduced models satisfying $\AE_A^x$ preserve \sATLs in the \iR semantics.

\begin{theorem}\label{prop:ae-atliR}
Let $A\subseteq\A$, $x \in \set{\Std,\React}$, and let models $\model' \subseteq \model$ satisfy $\AE_A^x$.
For each \sATLs formula $\varphi$, that refers only to coalitions $\hat{A}\subseteq A$,
we have that $\model,\iota \satisf[\iR] \varphi$\quad iff \quad $\model',\iota' \satisf[\iR] \varphi$.
\end{theorem}
\begin{proof}
Proof by induction on the structure of $\varphi$.
We show the case $\varphi = \coop{\hat{A}}\gamma$.
The cases for $\neg, \land$ are straightforward.

Notice that $\outcome_{\model'}^{x}(\iota,\strat_{\hat{A}}) \subseteq \outcome_{\model}^{x}(\iota,\strat_{\hat{A}})$,
which together with the condition $\AE_A^x$ implies that the sets $\outcome_{\model}^{x}(\iota,\strat_{\hat{A}})$ and $\outcome_{\model'}^{x}(\iota,\strat_{\hat{A}})$ are stuttering path equivalent.
Hence, the thesis follows from \Cref{equivs}.
\end{proof}

Together with \Cref{prop:stequ}, we obtain the following.
\begin{theorem}\label{prop:ae-atliR-corollary}
Let $\model$ be an I/O iCGS, and let $\model' \subseteq \model$ be the reduced model generated by
DFS with the choice of $E(\state')$ for $\state' \in \States'$ given by conditions {\bf C1, C2, C3}
and the independence relation $I_{A,\hatPV}$.
For each \sATLs formula $\varphi$ over $\hatPV$, that refers only to coalitions $\hat{A}\subseteq A$,
we have: $\model,\iota \satisf[\iR] \varphi$\quad iff \quad $\model',\iota' \satisf[\iR] \varphi$.
\end{theorem}

Thus, the reduction indeed preserves the satisfaction of $\ATL_\iR^*$ formulas without nested strategic modalities.